\documentclass{ecai} 

\usepackage{latexsym}
\usepackage{amssymb}
\usepackage{amsmath}
\usepackage{amsthm}
\usepackage{booktabs}
\usepackage{enumitem}
\usepackage{graphicx}
\usepackage{color}

\usepackage{xcolor}
\usepackage{booktabs}
\usepackage{multirow}
\usepackage{mathtools}
\usepackage{bbm}
\usepackage{dsfont}

\newcommand{\BibTeX}{B\kern-.05em{\sc i\kern-.025em b}\kern-.08em\TeX}

\begin{document}


\begin{frontmatter}


\paperid{123} 


\title{RaMen: Multi-Strategy Multi-Modal Learning \\ for Bundle Construction}





\author[A]{\fnms{Huy-Son}~\snm{Nguyen}\footnote{Shared first-authors.}}
\author[B]{\fnms{Quang-Huy}~\snm{Nguyen}\footnotemark}
\author[B]{\fnms{Duc-Hoang}~\snm{Pham}\footnotemark}
\author[B]{Duc-Trong Le}
\author[B]{Hoang-Quynh Le}
\author[C]{Padipat Sitkrongwong}
\author[C]{Atsuhiro Takasu}
\author[A]{Masoud Mansoury}


\address[A]{Delft University of Technology, The Netherlands}
\address[B]{VNU University of Engineering and Technology, Hanoi, Vietnam}
\address[C]{National Institute of Informatics, Japan}


\begin{abstract}
Existing studies on bundle construction have relied merely on user feedback via bipartite graphs or enhanced item representations using semantic information. These approaches fail to capture elaborate relations hidden in real-world bundle structures, resulting in suboptimal bundle representations. To overcome this limitation, we propose RaMen, a novel method that provides a holistic multi-strategy approach for bundle construction. RaMen utilizes both intrinsic (\emph{characteristics}) and extrinsic (\emph{collaborative signals}) information to model bundle structures through Explicit Strategy-aware Learning (\emph{ESL}) and Implicit Strategy-aware Learning (\emph{ISL}). ESL employs task-specific attention mechanisms to encode multi-modal data and direct collaborative relations between items, thereby explicitly capturing essential bundle features. Moreover, ISL computes hyperedge dependencies and hypergraph message passing to uncover shared latent intents among groups of items. Integrating diverse strategies enables RaMen to learn more comprehensive and robust bundle representations. Meanwhile, Multi-strategy Alignment \& Discrimination module is employed to facilitate knowledge transfer between learning strategies and ensure discrimination between items/bundles. Extensive experiments demonstrate the effectiveness of RaMen over state-of-the-art models on various domains, justifying valuable insights into complex item set problems.


\end{abstract}

\end{frontmatter}

\section{Introduction}

Bundle construction, which focuses on grouping relevant items into appealing offers, is an increasingly valuable strategy in marketing for both physical stores and e-commerce platforms across various industries~\cite{zhu2014bundle,sun2024revisiting,ma2024leveraging}. Beyond driving revenue growth, well-crafted bundles can enhance the customer experience by introducing variety and mitigating decision fatigue. 
Traditional bundle design has required manual effort from retailers, being not only time-consuming but also costly, making it difficult to scale across large datasets~\cite{sun2024revisiting}. The emergence of automatic bundle manufacturers has garnered attention from researchers due to its scalability and efficiency \cite{ma2024leveraging,sun2024revisiting}.

Most approaches oversimplify the complex strategies behind decision-making processes by relying on noisy datasets where bundles are defined using unreasonable heuristics~\cite{fang2018customized,sun2024revisiting,zhu2014bundle}. For example, some studies equate collaborative items with bundles without considering the context-specific nature of such combinations~\cite{liu2017modeling,deng2021build}. Others rely on user-generated lists in niche domains like \textit{music}~\cite{irene2019automatic} or \textit{gaming}~\cite{pathak2017generating}, limiting their general applicability. These studies often overlook the underlying rationale for bundling decisions, assuming that historical bundles are readily available for recommendation purposes, which is unrealistic in many real marketing scenarios.
State-of-the-art (SOTA) studies \cite{ma2024cirp,ma2024leveraging,du2023enhancing} on bundle-related tasks predominantly rely on user feedback, represented through bipartite graphs with LightGCN~\cite{he2020lightgcn}, or attempt to enhance item representations using semantic data~\cite{liu2025fine}. Yet, such approaches often lead to suboptimal bundle representations, making it difficult to accurately capture the underlying structure of real-world bundling strategies.

In practice, successful bundle creation hinges on leveraging the inherent relationships between products, enabling businesses to develop combinations that meet specific customer needs~\cite{sun2024revisiting}. 
To build potential bundles, it is vital to consider both intrinsic (\textit{characteristic}) and extrinsic (\textit{collaborative}) information of products, ensuring alignment with particular customer intents or preferences. Moreover, bundles are often tailored to target distinct customer groups or particular intentions, such as those curated by style, age, or price segmentation, etc.~\cite{sun2024revisiting}. 
Modeling shared latent attributes among items plays a significant role in determining optimal bundling tactics. 

{\textbf{Approaches and Contributions.}}  To address these limitations, we introduce RaMen, a novel framework that systematically models the bundle construction process through a multi-strategy multi-modal learning paradigm.
RaMen leverages both intrinsic (\textit{semantic}) and extrinsic (\textit{collaborative}) information to effectively capture the latent structure of bundles. This is achieved by incorporating two key components: Explicit Strategy-aware Learning and Implicit Strategy-aware Learning. The Explicit Strategy focuses on encoding essential bundle characteristics by utilizing task-specific attention mechanisms, which highlight direct item relationships and relevant semantic information as \textit{our first contribution}. Meanwhile, Implicit Strategy-aware Learning employs hypergraph message passing and hyperedge dependency matrices to uncover shared latent intents among item groups, capturing deeper implicit interactions that traditional models overlook as \textit{our second contribution}.
By integrating multi-strategy representations, RaMen constructs more comprehensive and generalizable bundle representations. Furthermore, Multi-strategy Alignment \& Dispersion is designed to enhance knowledge transfer between learning strategies while maintaining discrimination between different object representations. As \textit{our final contribution}, extensive experimental evaluations substantiate the efficacy of RaMen, revealing its ability to deliver novel insights and robust solutions to bundling problems.
To the best of our knowledge, this study is the first to model the collaborative relationships and characteristics of items, combined with learning shared attributes between them, to identify the hidden intents of each constructed bundle.

\section{Related Work}

With the rapid growth of e-commerce, studies related to complex item sets such as next-basket~\cite{li2023next,nguyen2023hhmc}, bundle recommendation~\cite{sun2024survey}, and bundle construction~\cite{ma2024leveraging} have garnered significant attention as a means to enhance business revenue and mitigate monotonous recommendations based on cross-selling concepts. 
While research on bundle recommendation focuses on suggesting pre-defined bundles based on user interactions~\cite{ma2022crosscbr,wei2023strategy,bui2024bridge,bui2025personalized}, our objective in bundle construction is to predict collections of items to create bundles that fulfill specific needs to attract users~\cite{ma2024leveraging,ma2024cirp}. 
Traditional bundle recommendation approaches rely on predefined criteria~\cite{zhu2014bundle} and matrix factorization \cite{pathak2017generating} to capture and leverage user preferences. Recent advancements in deep learning have significantly improved the performance of bundle recommendation systems, including attention-based techniques~\cite{chen2019matching}, graph neural networks~\cite{chang2021bundle,zhao2022multi}, contrastive learning~\cite{ma2022crosscbr,zhao2022multi}, generative methods~\cite{bui2024bridge,bui2025personalized} but still based purely on tripartite relations between user-bundle, bundle-item, user-item pairs.
Meanwhile, appropriately constructed bundles can enable the system to deliver more effective and targeted recommendations~\cite{sun2024revisiting,sun2024survey}.

Bundle construction tasks concentrate on completing partial bundles by identifying and selecting missing items from a pool of candidate products~\cite{ma2024leveraging,ma2024cirp,sun2024revisiting}. 
This process enables systems to automatically construct comprehensive and diverse bundles that better cater to a broad range of consumer preferences, ultimately enhancing product recommendations and improving user satisfaction. 
Common approaches in bundle construction leverage user-item interactions to uncover item-to-item relationships, thereby learning hidden bundle patterns to model the ultimate bundle representations~\cite{chang2021bundle,deng2021build,ding2023personalized}. 
Bundle representation learning consistently lies at the core of bundle-oriented challenges.
Sequential models, such as Bi-LSTM~\cite{han2017learning}, were employed to capture relations between consecutive items. 
However, as bundles are inherently unordered, conventional sequential models struggle to fully capture pairwise correlations. 
To tackle this issue, attention mechanisms~\cite{chen2019matching}, Transformers~\cite{wei2023strategy}, and graph neural networks (GNNs)~\cite{nguyen2024bundle} have been utilized to model both pairwise and higher-order item relationships. Despite these advances in item correlation modeling, limited attention has been paid to multimodal information, leading to construct bundles that lack coherence in their item characteristics as well as meaningless intents~\cite{sun2024revisiting}.

The integration of multi-modal data proves effective in addressing key challenges such as data sparsity and cold-start issues~\cite{liu2023multimodal,ma2024leveraging}. 
In the context of product bundling, several methods have leveraged multi-modalities to improve item representation learning. 
Recent SOTA model CLHE~\cite{ma2024leveraging} leverages self-attention mechanisms\cite{vaswani2017attention} to fuse multi-modal features with user feedback, focusing on addressing data sparsity and cold-start issues.  
However, CLHE~\cite{ma2024leveraging} improves the learning process by solely incorporating multi-modal features and a bipartite item-user graph with LightGCN~\cite{he2020lightgcn} that can not model rigorous relationships between anchor items and accessories to determine primary intents of bundles.
Another approach employs a multi-modal encoder along with cross-modal and cross-item contrastive loss to better capture item-to-item relationships \cite{ma2024cirp}. 
CIRP thrives on employing cross-item relation to provide the pre-training model of item representations \cite{ma2024cirp}. 
Furthermore, some promising approaches integrate large language models (LLMs) into the bundle construction process, enhancing the model's understanding of relationships between different modalities semantically~\cite{liu2025fine,sun2024adaptive}.
The available results are remarkable, but authoritative studies have not yet been able to effectively address the modeling of corporate strategies based on both of collaborative relationships and item characteristics. 

SOTA models on related tasks mostly rely on bipartite graphs with LightGCN via user feedbacks~\cite{ma2022crosscbr,ma2024leveraging,zhou2023tale,zhou2023enhancing,du2023enhancing}, or attempt to enhance item representations merely using semantic data~\cite{liu2025fine}. They often lead to suboptimal bundle representations, making it difficult to capture accurately the underlying structure of real-world bundling strategies.
Different from previous works~\cite{ma2024leveraging,sun2024survey,sun2024revisiting}, our multi-strategy multi-modal learning paradigm aims to to thoroughly model the collaborative relationships and characteristics of items, combined with learning shared attributes among items/bundles, to grasp the more comprehensive intents of each constructed bundle. 
Compared to the closest method CLHE~\cite{ma2024leveraging}, we inherit the design of their evaluation protocol and input-output flows because CLHE is the pioneer and SOTA research on multimodal bundle construction.
As mentioned above, CLHE solely incorporates multi-modal features and an item-user graph with LightGCN that can not model rigorous relationships between anchor items and accessories to determine primary intents of bundles like RaMen.
Meanwhile, ESL of RaMen not only encodes essential bundle characteristics by attention mechanisms, but also models association among items via our item-item graph. 
Notably, our refined attention-based propagation on item-item graph can learn more comprehensive associations among items, representing a significant improvement over the ubiquitous user-item LightGCN.
Moreover, to tackle the issues mentioned about multi-strategy bundle construction, our ISL is designed to uncover shared latent intents among item groups.
Besides the integration of two prime encoders, we devise MAD module to enhance knowledge transfer between learning strategies while maintaining discrimination between different items/bundles.

\section{Methodology}
\label{sec:method}
\begin{figure}[t!]
    \centering    
    \includegraphics[width=\linewidth]{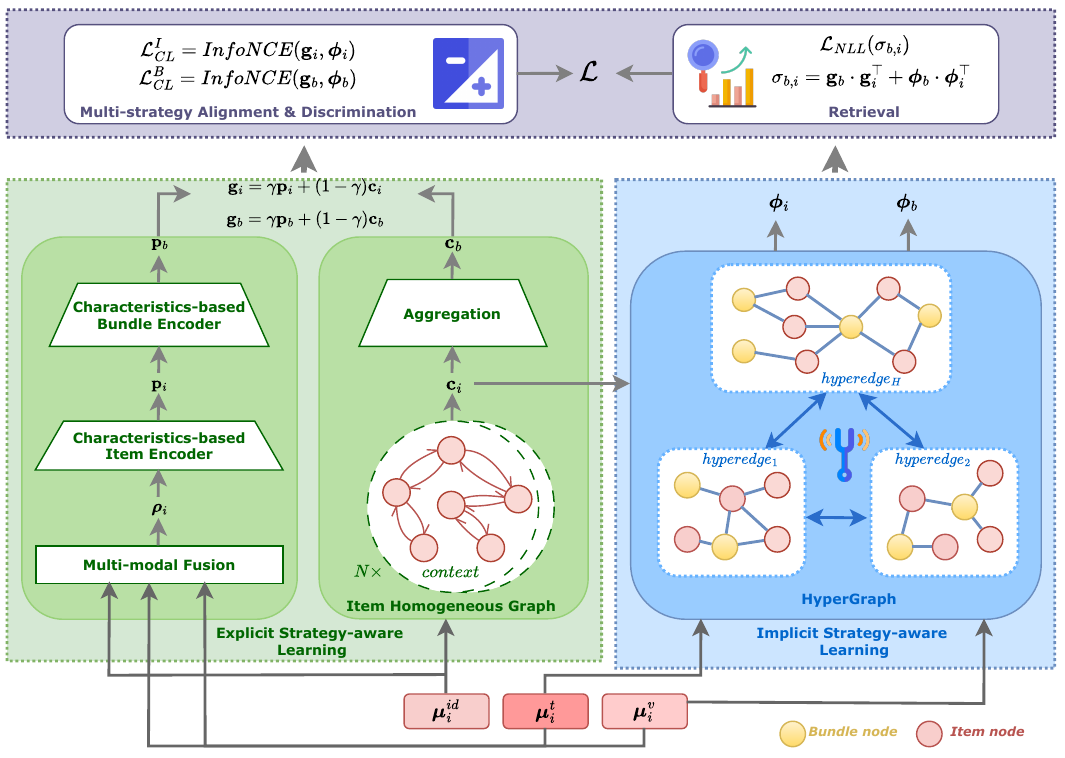}
    \vspace{-6mm}
    \caption{Overall architecture of our proposed model RaMen.}
    \label{fig:model}
\end{figure}

Section~\ref{sec:method} presents the overall architecture of RaMen as Fig.~\ref{fig:model} for bundle construction tasks, consisting of four main modules: (i) Explicit Strategy-aware Learning, (ii) Implicit Strategy-aware Learning, (iii) Multi-strategy Alignment \& Discrimination, and (iv) Retrieval \& Joint Optimization.


\subsection{Preliminaries}

\subsubsection{Problem Formulation.}

For the tasks of bundle construction, let $\mathcal{I} = \{ i_k \}_{k=1}^{|\mathcal{I}|}$,  $\mathcal{U} = \{u_k \}_{k=1}^{|\mathcal{U}|}$, $\mathcal{B} = \{b_k \}_{k=1}^{|\mathcal{B}|}$ represents a set of items, users, and bundles, respectively.
Relied on historical user behaviors, the user-item interaction is collected formally as a binary matrix $X \in \{0,1\}^{|\mathcal{U}| \times |\mathcal{I}|}$, where $X_{u,i} = 1$ indicates user $u$ interacted with item $i$, and $X_{u,i} = 0$ otherwise.
Likewise, the bundle-item affiliation is defined in matrix $Y \in \{0,1\}^{|\mathcal{B}| \times |\mathcal{I}|}$, where $Y_{b,i} = 1$ if bundle $b$ contains item $i$, and $Y_{b,i}=0$ otherwise. In particular, each bundle $b = \{ i_h \}_{h=1}^{|b|} \in \mathcal{B}$ is a collection of pertinent items. 
We establish the training set of bundles as $\widetilde{\mathcal{B}}=\{ b_k \}_{k=1}^{|\widetilde{\mathcal{B}}|} \subset \mathcal{B}$, and testing set of bundles as $\widehat{\mathcal{B}}=\{ b_k \}_{k=|\widetilde{\mathcal{B}}|+1}^{|\mathcal{B}|} \subset \mathcal{B}$.
Given partial bundle $\hat{b}$ containing a few seed items from each unseen testing bundle ${b} \in \widehat{\mathcal{B}}$, the objective of bundle construction is to efficiently predict the deficient items $\{{b} \setminus \hat{b}\}$ to capture the comprehensive bundle. 
The training process complies with an auto-encoder approach~\cite{ma2024leveraging}, where the entire items within the bundle are considered as the input, and the same set should be predicted as the output.

\subsubsection{Semantic Information Extraction.}
Inherited from \cite{liu2023multimodal,ma2024leveraging}, the textual and visual features of each item are derived from large-scale multi-modal feature extractors, represented as $\{\mathbf{m}_i^t \in \mathbb{R}^{d_{mt}}, \mathbf{m}_i^v \in \mathbb{R}^{d_{mv}}\}$, where $d_{mt}$ and $d_{mv}$ denote the dimensions of the textual and visual embeddings. Textual information, such as the item's title and description, is encoded into $\mathbf{m}_i^t$, while images are encoded into $\mathbf{m}_i^v$.
The encoded visual and textual embeddings are respectively transformed into a unified latent space via specialized refinement MLP networks such as $\textbf{MLP}^v, \textbf{MLP}^t$ to mitigate the misalignment caused by dimensional differences across modalities \cite{liu2022disentangled,liu2023multimodal}. This process is derived as follows:
\begin{equation}
    \boldsymbol{\mu}_i^v = \text{\textbf{MLP}}^v (\mathbf{m}_i^v), \quad
    \boldsymbol{\mu}_i^t = \text{\textbf{MLP}}^t (\mathbf{m}_i^t),
\end{equation}
$\boldsymbol{\mu}_i^v, \boldsymbol{\mu}_i^t \in \mathbb{R}^d$ are aligned embeddings after dimension adjustment.


\subsubsection{Item-level Collaborative Relation Construction.}
Learning collaborative signals based on user-item graphs~\cite{liu2023multimodal,ma2024leveraging} may introduce noise when propagating higher-order collaborative signals. To tackle this limitation, an item homogeneous graph $\mathcal{G}_{\mathcal{I}} = \{\mathcal{I}, \mathcal{E}_{\mathcal{I}}\}$ is designed to learn direct influences among items,  where ${\mathcal{I}}$ and $\mathcal{E}_{\mathcal{I}} = \{e_{i,j} | i,j \in \ \mathcal{I} \}$  represent the set of vertices and edges. An item co-purchased matrix $E \in \mathbb{R}^{|\mathcal{I}| \times |\mathcal{I}|}$ is first computed, where $E = X^\top \cdot X$. The direct relation $e_{i,j}$ between each pair of items $i, j$ is established by discretizing the weighted matrix $E$ with threshold $\epsilon$ into an unweighted version to facilitate information propagation, derived as $e_{i,j}=1 ~\text{if } E_{i,j} \geq \epsilon \land  i \neq j$, and $0$ otherwise\footnote{This work can be developed more robustly by adaptively filtering noisy edges in the graph instead of empirical selection across diverse domains.}. 

\subsection{Explicit Strategy-aware Learning}


\subsubsection{Characteristic Strategy Encoder.}
Given the assumption that bundle construction is driven by enterprises leveraging distinctive characteristics~\cite{sun2024revisiting,zhu2014bundle}, this module aims to synthesize relevant multi-modal features across items within each bundle, optimizing the bundling strategy from a characteristic-based perspective. 

\textbf{Multi-modal Fusion}.
The obtained semantic embeddings $\boldsymbol{\mu}_i^v, \boldsymbol{\mu}_i^t$, along with the initialized ID embedding $\boldsymbol{\mu}_i^{id}$ in the same latent space $\mathbb{R}^d$ are synthesized into multi-modal item representation $\boldsymbol{\rho}_i \in\mathbb{R}^{2\times d}$ through concatenation-based fusion, mathematically represented as:   

\begin{equation}
    \boldsymbol{\rho}_i = \Xi ( \mathbf{\Psi}_\rho ~(\boldsymbol{\mu}_i^v ~\| ~\boldsymbol{\mu}_i^t), ~\boldsymbol{\mu}_i^{id} ~),
\end{equation}
where $\mathbf{\Psi}\rho \in \mathbb{R}^{d \times 2d}$ is the linear transformation matrix, and $||$ denotes the vertical concatenation of the semantic features. Besides, $\Xi$ performs horizontal concatenation, synthesizing these components into the multi-modal feature matrix of items.

\textbf{Characteristics-based Item Encoder}. Leveraging the proven effectiveness of attention mechanisms~\cite{vaswani2017attention} and the diverse aspects of item features in recommender systems~\cite{deng2023multi,liu2023multimodal}, RaMen employs self-attention techniques to compute correlation scores between item-level multi-modal characteristics, formalized as:

\begin{equation}
    \boldsymbol{\tilde{\rho}}_i^{(l)} = \text{softmax}\underbrace{\left(\frac{1}{\sqrt{d}} \boldsymbol{\tilde{\rho}}_i^{(l-1)} \mathbf{\Psi}_I^K \left(\boldsymbol{\tilde{\rho}}_i^{(l-1)} \mathbf{\Psi}_I^Q\right)^\top\right)}_{\text{\textit{correlation score of item characteristics}}}   \boldsymbol{\tilde{\rho}}_i^{(l-1)},
\label{eq:item_attention}
\end{equation}
where $\mathbf{\Psi}_I^K$ and $\mathbf{\Psi}_I^Q \in \mathbb{R}^{d \times d}$ represent the key and query projection matrices. The feature matrix at layer $l$-th, denoted as $\boldsymbol{\tilde{\rho}}_i^{(l)} \in \mathbb{R}^{2 \times d}$, evolves from the initial features $\boldsymbol{\tilde{\rho}}_i^{(0)} = \boldsymbol{{\rho}}_i$. 
The feature matrix $\boldsymbol{\tilde{\rho}}_i^{(L_1)} \in \mathbb{R}^{2 \times d}$ of item $i$ is produced after $L_1$ attention layers, and the corresponding characteristic vector $\mathbf{p}_i \in \mathbb{R}^d$ is subsequently computed by mean pooling over the feature matrix $\boldsymbol{\tilde{\rho}}_i^{(L_1)}$.



\textbf{Characteristics-based Bundle Encoder}. RaMen is capable of capturing the critical semantic features of items, enhancing the bundle construction process by focusing on the intricate correlations between multi-modal characteristics. With the obtained item embeddings, the bundle characteristics is formed by concatenating the embeddings of components, expressed as $\boldsymbol{\rho}_b = \Xi (\{\mathbf{p}_i\}_{i \in b})$.
The bundle representation is refined through $L_2$ attention layers, defined as:

\begin{equation}
\begin{aligned}
        \boldsymbol{\tilde{\rho}}_b^{(l)} = \text{softmax}\underbrace{\left(\frac{1}{\sqrt{d}} \boldsymbol{\tilde{\rho}}_b^{(l-1)} \mathbf{\Psi}_B^K \left(\boldsymbol{\tilde{\rho}}_i^{(l-1)} \mathbf{\Psi}_B^Q\right)^\top\right)}_{\text{\textit{correlation score of bundle characteristics}}}   \boldsymbol{\tilde{\rho}}_b^{(l-1)},
\end{aligned}
\label{eq:bundle_attention}
\end{equation}
where $\boldsymbol{\tilde{\rho}}_b^{(l)}$ denotes the feature matrix of bundle $b$ in layer $l$-th with the initial value $\boldsymbol{\tilde{\rho}}_b^{(0)}=\boldsymbol{\rho}_b$; $\mathbf{\Psi}_B^K, \mathbf{\Psi}_B^Q\in \mathbb{R}^{d\times d}$ represent the learnable projection matrices. After refining through $L_2$ attention layers, we adopt mean pooling to each bundle feature matrix $\boldsymbol{\tilde{\rho}}_b^{(L_2)}$, aggregating the ultimate bundle characteristic $\mathbf{p}_b\in\mathbb{R}^d$ correspondingly.

\subsubsection{Collaborative Strategy Encoder.}

Given the assumption that an effective bundle construction strategy should leverage item-level collaborative relations to align logically with user expectations, Collaborative Strategy Encoder employs advanced attention mechanisms \cite{brodyattentive,nguyen2024bundle} to effectively propagate high-order collaborative signals with weighted causal influences among nodes of graph $\mathcal{G}_{\mathcal{I}}$ underlying various contexts. The propagation of item neighborhood features is derived as:

\begin{equation}
    \alpha_{i \leftarrow j}^{(n)} = \frac{\exp \left( \mathbf{q}_{(n)}^\top  \varphi (\mathbf{\Psi}^{(n)}  \mathbf{s}_i + \mathbf{\hat{\Psi}}^{(n)}  \mathbf{s}_j + \mathbf{\Delta}) \right) }{\sum_{j' \in \mathcal{N}_{i}} \exp \left( \mathbf{q}_{(n)}^\top \varphi(\mathbf{\Psi}^{(n)}  \mathbf{s}_i + \mathbf{\hat{\Psi}}^{(n)} \mathbf{s}_{j'} + \mathbf{\Delta}) \right) }  \label{eq:asym}
\end{equation} 
where $\mathbf{q}_{(n)}\in\mathbb{R}^d$ is a learnable context vector; $\mathbf{s}_i, \mathbf{s}_j \in\mathbb{R}^d$ represent the embeddings of item $i$ and $j$; $\mathbf{\Delta}$ denotes the bias weights; and $\mathcal{N}_i$ denotes neighborhood set of item $i$. The symbol $\varphi$ presents the activation function LeakyReLU.
Compared to conventional attention techniques in GNN~\cite{gao2023survey}, we employ specialized transformation matrices $\mathbf{\Psi}^{(n)}, \mathbf{\hat{\Psi}}^{(n)} \in \mathbb{R}^{d \times d}$ for the target-item node $i$ and source-item node $j$ at the $n$-th context to mitigate overfitting.
The ID embeddings of items are adopted as input to Collaborative Strategy Encoder.


The final representation $\mathbf{c}_i\in\mathbb{R}^d$ of item $i$ is aggregated after $N$ contexts, derived as follows:
\begin{equation}
\begin{aligned}
   \mathbf{s}_i^{(n)} &= \sum_{j\in\mathcal{N}_i} \alpha^{(n)}_{i \leftarrow j} \mathbf{\Psi}_{j}^{(n)} \mathbf{s}_j ,\\
\mathbf{c}_i &= \beta\frac{1}{N}\sum_{n=1}^N \mathbf{s}_i^{(n)} + (1-\beta)\mathbf{s}_i
\end{aligned}
\label{eq:coll}
\end{equation}
where $\mathbf{s}_i^{(n)} \in\mathbb{R}^d$ signifies the latent representation of item $i$ at the $n$-th layer, $\beta$ modulates the impact of the residual connection on the enhanced item embedding.
Thereby, the Collaborative Strategy Encoder obtains the bundle embedding $\mathbf{c}_b \in \mathbb{R}^d$ through the mean aggregation of the embeddings of items within bundle $b$. 



The obtained representations of items/bundles from Characteristic Strategy Encoder and Collaborative Strategy Encoder are aggregated to compute Explicit Strategy-aware embeddings $\mathbf{g}_b$ and $\mathbf{g}_i$ for bundle $b$ and item $i$, derived as follows:
\begin{align}
    \mathbf{g}_i &= \gamma \mathbf{p}_i +  (1-\gamma) \mathbf{c}_i, \\
    \mathbf{g}_b &= \gamma \mathbf{p}_b + (1-\gamma)  \mathbf{c}_b,
\end{align}
where $\gamma$ controls the effect of embeddings from different encoders.

\subsection{Implicit Strategy-aware Learning}

The hypergraph architecture \cite{guo2024lgmrec,gao2023survey,xia2022hypergraph}, which extends beyond-pairwise relations, enables the latent representation of both intra-bundle and inter-bundle relations by modeling shared attributes among items as hyperedges.
To effectively capture implicit strategies within groups of items, we introduce learnable hyperedge embeddings $W_m \in \mathbb{R}^{H \times d}$, designed to encode latent attributes specific to each modality $m \in \{t,v\}$, where $H$ represents the number of hyperedges and $t,v$ respectively denote textual/visual features. The dependency matrices for hyperedges and items/bundles are formally constructed as follows:
\begin{equation}
    F_{\mathcal{I}}^m = M_\mathcal{I}^m ~ (~W_m~)^\top, \quad
    F_{\mathcal{B}}^m = Y~(~F_{\mathcal{I}}^m~)^\top,
\end{equation}
where $F_\mathcal{I}^m \in \mathbb{R}^{|\mathcal{I}|\times H}$ and $F_\mathcal{B}^m\in\mathbb{R}^{|\mathcal{B}|\times H}$ are item-hyperedge and bundle-hyperedge dependency matrices, respectively; $M_\mathcal{I}^m = \{ \boldsymbol{\mu}_i^m \}_{i \in \mathcal{I}}$ is the feature matrix of modality $m$. 
The matrix $F_\mathcal{I}^m$ aims to capture the connections between items and hyperedges, grouping similar items under shared attributes. Besides, the matrix $F_\mathcal{B}^m$ reflects how bundles are indirectly associated with hyperedges via the items they contain. The stronger the affiliation between a bundle and items linked to latent attributes, the more likely the bundle’s strategy is aligned with that attribute.
Inspired by~\cite{xia2022hypergraph,guo2024lgmrec}, Gumbel-Softmax reparameterization technique \cite{jang2017categorical} is adopted to mitigate the impact of noisy connections between items/bundles and hyperedges, defined as follows:

\begin{equation}
\mathbf{\hat{f}}^m_{i} = \text{softmax} \left( \frac{\log \boldsymbol{\theta} - \log (1 - \boldsymbol{\theta}) + \mathbf{f}^m_{i}}{\tau} \right),
\end{equation}
where $\mathbf{\hat{f}}_i^m \in \mathbb{R}^H$ represents the relation vector of item $i$ with hyperedges in the fine-grained dependency matrix $\hat{F}_\mathcal{I}^m$; each value of the noise vector $\boldsymbol{\theta}$ is sampled from a uniform distribution in range $[0,1]$; and temperature parameter $\tau$ is empirically selected as $0.2$. Likewise, we obtain the fine-grained bundle-hyperedge matrix $\hat{F}_b^m$. These fine-grained dependency matrices are then leveraged to propagate item and bundle attributes relied on each modality, derived as:

\begin{equation}
\begin{aligned}
\boldsymbol{\phi}_i^{m,(z+1)} = \hat{F}_\mathcal{I}^m \cdot (\hat{F}_\mathcal{I}^{m})^{\top}\cdot \boldsymbol{\phi}_i^{m,(z)}, \\
\boldsymbol{\phi}_b^{m,(z+1)} = \hat{F}_\mathcal{B}^m \cdot (\hat{F}_\mathcal{I}^m)^{\top}  \cdot \boldsymbol{\phi}_i^{m,(z)}, 
\end{aligned}
\end{equation}
where $\boldsymbol{\phi}_i^{m,(z)}$ is the embedding of item $i$ corresponding to modality $m$ at the $z$-th hypergraph layer, 
specifically $\boldsymbol{\phi}_i^{m,(0)}=\mathbf{c}_i$. The final representations of Implicit Strategy-aware Learning are obtained after propagating $Z$ hypergraph layers as follows:

\begin{equation}
     \boldsymbol{\phi}_i = \varPsi\Bigg(\sum_{m\in \{v,t\}}  \boldsymbol{\phi}_i^{m, (Z)}\Bigg),\quad  
     \boldsymbol{\phi}_b = \varPsi\Bigg(\sum_{m\in \{v,t\}}  \boldsymbol{\phi}_b^{m, (Z)}\Bigg)\,
\end{equation}
where $\varPsi$ is $L_p$ normalization function, $\boldsymbol{\phi}_i^{m,(Z)}$ and $\boldsymbol{\phi}_b^{m,(Z)}\in\mathbb{R}^d$ are embeddings of item $i$ and bundle $b$ corresponding to modality $m$ obtained after $Z$ hypergraph layers, respectively.

\subsection{Multi-strategy Alignment \& Discrimination}

We apply the contrastive loss, specifically InfoNCE \cite{oord2018representation}, to align the representations of the same item or bundle generated by different strategies and ensure the separation of embeddings corresponding to distinct items or bundles within the embedding space. This technique leads to more coherent and discriminative representation for each item or bundle. The item-level contrastive loss for multi-strategy learning is derived as:

\begin{equation}
    \mathcal{L}_{CL}^{I} = \frac{1}{|\mathcal{I}|} \sum_{i \in \mathcal{I}} - \log \frac{\exp \left( \cos(\mathbf{g}_i, \boldsymbol{\phi}_i) / \tau \right))}{\sum_{j \in \mathcal{I}} \exp\left(\cos(\mathbf{g}_i, \boldsymbol{\phi}_j) / \tau \right) },
\end{equation}
where $\text{cos}(\cdot)$ performs cosine similarity function. The bundle-level contrastive loss $\mathcal{L}_{CL}^{B}$ is derived similarly. The objective loss of Multi-strategy Alignment \& Discrimination module is to reconcile
different strategy-based representations as:

\begin{equation}
    \mathcal{L}_{MAD} = \mathcal{L}_{CL}^{I} +  \mathcal{L}_{CL}^{B},
\end{equation}

\subsection{Retrieval \& Joint Optimization}

\subsubsection{Retrieval.}
To estimate the possibility that item $i$ belongs to bundle $b$, we adopt the inner product to compute score $\sigma_{b,i}$ from multi-strategy representations, as follows:

\begin{equation}
    \sigma_{b,i} = \mathbf{g}_b \cdot \mathbf{g}_i^\top + \boldsymbol{\phi}_b\cdot \boldsymbol{\phi}_i^\top
\end{equation}
  
Ground in the steering study of \cite{ma2024leveraging}, the negative log-likelihood (NLL) is employed as the primary optimization objective after obtaining the score $\sigma_{b,i}$. By using NLL loss, the model learns to assign higher scores to items that are likely to belong to a bundle while minimizing scores for irrelevant items. Here, the NLL loss for optimizing prediction is defined as:
\begin{equation}
    \mathcal{L}_{NLL} = \frac{1}{|\mathcal{\tilde{B}}|}\sum_{b\in\mathcal{\tilde{B}}}\frac{1}{|\mathcal{I}|}\sum_{i\in\mathcal{I}} -\mathds{1}_{i\in b}\: \text{log} \left(\frac{\text{exp}(\sigma_{b,i})}{\sum_{j\in\mathcal{I}}\text{exp}(\sigma_{b,j})} \right),
\end{equation}
where $\mathds{1}_{i\in b}$ represents an indicator function that equals 1 if the component item $i$ belongs to the bundle $b$, and 0 otherwise.

\subsubsection{Joint Optimization.} 
The overall objective function $\mathcal{L}$ is composed of the defined loss functions combined with a regularization term, formulated as:
\begin{equation}
    \mathcal{L} = \mathcal{L}_{NLL} + \lambda_1 \mathcal{L}_{MAD} + \lambda_2 \| \mathbf{\Theta} \|_2^2,
\end{equation}
where the hyperparameter $\lambda_1$ controls the impact of contrastive-based loss, and $\lambda_2$ denotes regularization weight with all the trainable parameters $\mathbf{\Theta}$ of model.

\section{Experiments}
We conduct extensive experiments to evaluate the effectiveness of RaMen, and analyze the significance of its main components. Moreover, some qualitative showcases accentuate the superior performance of RaMen compared to CLHE.
Our repository is available on Github to facilitate reproducibility and extension.


\subsection{Experimental Settings}
\subsubsection{Datasets and Evaluation Protocols.}

We utilize four datasets in diverse domains \cite{chen2018recsys,chen2019pog,sun2024revisiting}, as detailed in Table \ref{tab:dataset_stats}. POG~\cite{chen2019pog} considers fashion outfits as bundles, and many music tracks in the same session of Spotify  \cite{chen2018recsys} are treated as bundles. 
Bundles of Food and Electronic~\cite{sun2024revisiting} are constructed with high-quality metadata and meticulous intents.
Pre-trained BLIP~\cite{li2022blip} is adopted to extract visual and textual embeddings across these datasets.
To make fair comparisons with baselines, this work inherits the features extracted from baseline work~\cite{ma2024leveraging} for Spotify and POG. We split all bundles into train:valid:test set with a ratio of $7:1:2$ for four datasets. Within the valid set and test sets, items in each bundle are randomly masked as the target items to be predicted, while the remaining items form the partial bundle~\cite{ma2024leveraging}. The ubiquitous retrieval metrics \cite{gao2023survey,liu2023multimodal,ma2024leveraging,sun2024revisiting}, such as Recall@K ($R@K$) and NDCG@K ($N@K$), are employed to evaluate the prediction of models.

\begin{table}[t!]
    \centering
    \caption{The statistics of four benchmark datasets in diverse domains for bundle construction.}
    \resizebox{1\columnwidth}{!}{
    \begin{tabular}{lcccccccccc}
        \toprule
        \textbf{Dataset} & \textbf{\#U} & \textbf{\#I} & \textbf{\#B} & \textbf{\#B-I} & \textbf{\#U-I} & \textbf{Avg.I/B}  & \textbf{Avg.I/U} & \textbf{U-I Dens.} \\
        \toprule
        \textbf{POG} & 17,449 & 48,676 & 20,000 & 72,224 & 237,519 & 3.61  & 13.61 & 0.0073\% \\
        \textbf{Spotify} & 118,994 & 254,155 & 20,000 & 1,268,716 & 36,244,806 & 63.44 & 304.59  & 0.1198\% \\
        \textbf{Electronic} & 888 & 3,499 & 1,750 & 6,165 & 6,165 & 3.52  & 6.94  & 0.1984\% \\
        \textbf{Food} & 879 & 3,767 & 1,784 & 6,395 & 6,395 & 3.58 & 7.28  & 0.1931\% \\
        \bottomrule
    \end{tabular}
    }
    \label{tab:dataset_stats}
\end{table}

\subsubsection{Comparative Baselines.}
Based on the groundbreaking work in bundle construction task~\cite{ma2024leveraging}, we take into account the following baselines\footnote{Due to space limitation, `Trans' is an abbreviation for the `Transformer'-based model.}:
\textbf{Bi-LSTM} \cite{han2017learning},
\textbf{HyperGraph}~\cite{yu2022unifying},
\textbf{Trans} \cite{wei2023strategy},
\textbf{TransCL} \cite{ma2024leveraging},
\textbf{GAT} \cite{brodyattentive},
\textbf{CLHE}~\cite{ma2024leveraging}. 
In this study, GAT utilizes a graph attention mechanism to propagate high-order bundle-item affiliations, then computes the ultimate prediction as other comparative models. Meanwhile, the other baselines  are followed to the experimental setups of~\citet{ma2024leveraging}.



\subsubsection{Implementation Details.}
According to related works~\cite{ma2022crosscbr,ma2024leveraging,nguyen2024bundle,sun2024revisiting}, RaMen adopts Xavier initialization 
and Adam optimizer~\cite{kingma2014adam}, setting the prevalent configuration including the embedding size as $64$, the batch size as $1024$, the learning rate as $1e-3$ and regularization weight as $1e-5$. 
The hyperparameters are tuned by empirical studies, according to the related studies we inherit for each module. Inherited~\cite{zhou2023enhancing,zhou2023tale}, $\epsilon$ is tuned in increments based on dataset size. 
The values of $L_1, L_2, N, Z$ are empirically explored within range $\{1, 2, 3, 4, 5\}$,
and $\epsilon$ is set as $5$ for POG, $450$ for Spotify, $1$ for Food/Electronic relied on its interaction distribution. 
The number of hyperedges $H$ is chosen across $\{4, 8, 16, 32, 64\}$, and $\beta, \gamma, \lambda_1$ are tuned in range $\{0.1, 0.2, \dots, 0.8, 0.9\}$. 
RaMen is implemented using PyTorch, and trained on NVIDIA A100 80GB GPUs \& T4 15GB GPUs.
Baselines are conducted in the same configuration and acknowledged available results in the steering work by \citet{ma2024leveraging}. Bi-LSTM results should be merely acknowledged according to \cite{ma2024leveraging}, as we could not reproduce the same performance on Spotify.
Our repository is available on Github via  \texttt{\url{https://github.com/Rec4Fun/RaMen}}.


\begin{table*}[!ht]
\renewcommand{\arraystretch}{1}
\centering
\resizebox{0.7\textwidth}{!}{
    \begin{tabular}{c|c|cccccc|c|c}
    \toprule
    \multicolumn{1}{c|}{\textbf{Dataset}} & \multicolumn{1}{c|}{\textbf{Metric}} & \multicolumn{1}{c}{Bi-LSTM} & \multicolumn{1}{c}{HyperGraph} & \multicolumn{1}{c}{Trans} & \multicolumn{1}{c}{TransCL} & \multicolumn{1}{c}{GAT} & \multicolumn{1}{c|}{CLHE} & \multicolumn{1}{c|}{\textbf{RaMen}} & \multicolumn{1}{c}{$\%$ $\color{teal}\uparrow$} \\ 
    \midrule
    \multirow{4}{*}{\textbf{POG}} & 
    \textbf{R@10} & 0.0101 & 0.0113 & 0.0145 & 0.0160 & 0.0144 & \underline{0.0213} & $\textbf{0.0264}^\ddagger$ & 23.94 \\
    & \textbf{N@10} & 0.0072 & 0.0074 & 0.0097 & 0.0109 & 0.0098 & \underline{0.0160} & $\textbf{0.0191}^\ddagger$ & 19.38 \\
    & \textbf{R@20} & 0.0170 & 0.0207 & 0.0215 & 0.0202 & 0.0208 & \underline{0.0284} & $\textbf{0.0375}^\ddagger$ & 32.04 \\
    & \textbf{N@20} & 0.0097 & 0.0111 & 0.0114 & 0.0134 & 0.0118 & \underline{0.0193} & $\textbf{0.0226}^\ddagger$ & 17.10 \\
    \midrule
    \multirow{4}{*}{\textbf{Spotify}} & 
    \textbf{R@10} & - & 0.0306 & 0.0552 & 0.0593 & 0.0506 & \underline{0.0689} & \textbf{0.0695} & 0.87 \\
    & \textbf{N@10} & - & 0.0923 & 0.1587 & 0.1698 & 0.1493 & \underline{0.1950} & $\textbf{0.2060}^\ddagger$ & 5.64 \\
    & \textbf{R@20} & 0.0833 & 0.0572 & 0.0875 & 0.1014 & 0.0824 & \underline{0.1081} & \textbf{0.1091} & 0.83 \\
    & \textbf{N@20} & 0.1486 & 0.0941 & 0.1460 & 0.1696 & 0.1390 & \underline{0.1806} & $\textbf{0.1882}^\ddagger$ & 4.21 \\
    \midrule
    \multirow{4}{*}{\textbf{Electronic}} & 
    \textbf{R@10} & 0.0352 & 0.0616 & 0.1952 & 0.2355 & 0.3536 & \underline{0.4407} & $\textbf{0.7410}^\ddagger$ & 68.14 \\
    & \textbf{N@10} & 0.0242 & 0.0344 & 0.1294 & 0.1562 & 0.2643 & \underline{0.3300} & $\textbf{0.5104}^\ddagger$ & 54.67 \\
    & \textbf{R@20} & 0.0574 & 0.0928 & 0.2555 & 0.3050 & 0.3943 & \underline{0.4721} & $\textbf{0.8371}^\ddagger$ & 77.31 \\
    & \textbf{N@20} & 0.0298 & 0.0430 & 0.1456 & 0.1757 & 0.2812 & \underline{0.3390} & $\textbf{0.5373}^\ddagger$ & 58.50 \\
    \midrule
    \multirow{4}{*}{\textbf{Food}} & 
    \textbf{R@10} & 0.0189 & 0.0712 & 0.2453 & 0.2346 & 0.3793 & \underline{0.4557} & $\textbf{0.7575}^\ddagger$ & 66.23 \\
    & \textbf{N@10} & 0.0071 & 0.0379 & 0.1783 & 0.1769 & 0.2806 & \underline{0.3237} & $\textbf{0.5028}^\ddagger$ & 55.33 \\
    & \textbf{R@20} & 0.0350 & 0.1055 & 0.3137 & 0.3088 & 0.4097 & \underline{0.5077} & $\textbf{0.8459}^\ddagger$ & 66.61 \\
    & \textbf{N@20} & 0.0114 & 0.0478 & 0.1983 & 0.1985 & 0.2917 & \underline{0.3386} & $\textbf{0.5242}^\ddagger$ & 54.81 \\
    \bottomrule
    \end{tabular}
    }

\caption{Overall performances of RaMen compared with competitive baselines on four benchmark datasets from diverse domains.
The best results are in \textbf{bold}, and the second best results are \underline{underlined}. The symbol $\ddagger$ indicates statistically significant improvements over the second-best models with $p-value < 0.05$ obtained through the average performance of five runs of each model.
}
\label{tab:compare_table}
\end{table*}

\vspace{-4mm}
\subsection{Performance Comparison}
Table~\ref{tab:compare_table} demonstrates RaMen's effectiveness and adaptability in various scenarios with different domains and distributions, proving its superior performance in bundle construction compared to SOTA approaches.
The most significant improvements are indicated on benchmark datasets with small-sized bundle structures targeting specific intents, where RaMen achieves up to $77.31\%$, $66.61\%$, and $32.04\%$ higher \textit{w.r.t} $R@20$ compared to the strongest baseline CLHE on Electronic, Food, and POG, respectively. 
Besides, RaMen significantly outperforms all competitive attention-based architectures, such as Trans, TransCL, GAT and CLHE, exemplifying its ability to comprehensively encode essential features hidden in both characteristic and collaborative strategies.
Despite dealing with noise in learning target strategies for large bundles caused by numerous high-impact items within each bundle~\cite{nguyen2024bundle,wei2023strategy}, RaMen maintains modest yet steady gains on Spotify, especially in terms of ranking performance.
We attribute this robustness to RaMen's ability to dexterously model distinct decision-making strategies while integrating MAD module to enhance knowledge exchange between them.
These observations prove RaMen adeptly exploits both explicit and implicit strategies, combining their potential to facilitate optimal decisions in this task.

\begin{table*}[!t]
\renewcommand{\arraystretch}{1}
\centering
\resizebox{0.8\textwidth}{!}{
\begin{tabular}{c|c|cccccc}
\toprule
\multicolumn{1}{c|}{\textbf{Dataset}} & \multicolumn{1}{c|}{\textbf{Metric}} & \multicolumn{1}{c}{\textit{w/o CrSE}} & \multicolumn{1}{c}{\textit{w/o CbSE}} & \multicolumn{1}{c}{\textit{w/o ISL}} & \multicolumn{1}{c}{\textit{w/o MAD}} &  \multicolumn{1}{c}{\textit{w/o V}} & \multicolumn{1}{c}{\textit{w/o T}} \\ 
\midrule
\multirow{2}{*}{\textbf{POG}} 
& \textbf{R@20} & 0.0218$_{(\color{teal}\downarrow 41.87)}$ & 0.0301$_{(\color{teal}\downarrow 19.73)}$ & 0.0335$_{(\color{teal}\downarrow 10.67)}$ & 0.0348$_{(\color{teal}\downarrow 7.20)}$ & 0.0332$_{(\color{teal}\downarrow 11.47)}$ & 0.0293$_{(\color{teal}\downarrow 21.87)}$ \\
& \textbf{N@20} & 0.0119$_{(\color{teal}\downarrow 47.35)}$ & 0.0209$_{(\color{teal}\downarrow 7.52)}$ & 0.0202$_{(\color{teal}\downarrow 10.62)}$ & 0.0210$_{(\color{teal}\downarrow 7.08)}$ & 0.0204$_{(\color{teal}\downarrow 9.73)}$ & 0.0171$_{(\color{teal}\downarrow 24.34)}$ \\
\midrule
\multirow{2}{*}{\textbf{Spotify}} 
& \textbf{R@20} & 0.1028$_{(\color{teal}\downarrow 5.77)}$ & 0.0995$_{(\color{teal}\downarrow 8.80)}$ & 0.1071$_{(\color{teal}\downarrow 1.83)}$ & 0.1080$_{(\color{teal}\downarrow 1.01)}$ & 0.1041$_{(\color{teal}\downarrow 4.58)}$ & 0.1055$_{(\color{teal}\downarrow 3.30)}$ \\
& \textbf{N@20} & 0.1772$_{(\color{teal}\downarrow 5.84)}$ & 0.1662$_{(\color{teal}\downarrow 11.69)}$ & 0.1863$_{(\color{teal}\downarrow 1.01)}$ & 0.1838$_{(\color{teal}\downarrow 2.34)}$ & 0.1806$_{(\color{teal}\downarrow 4.04)}$ & 0.1830$_{(\color{teal}\downarrow 2.76)}$ \\
\midrule
\multirow{2}{*}{\textbf{Electronic}} 
& \textbf{R@20} & 0.7376$_{(\color{teal}\downarrow 11.89)}$ & 0.3560$_{(\color{teal}\downarrow 57.47)}$ & 0.8081$_{(\color{teal}\downarrow 3.46)}$ & 0.7836$_{(\color{teal}\downarrow 6.39)}$ & 0.7383$_{(\color{teal}\downarrow 11.80)}$ & 0.7516$_{(\color{teal}\downarrow 10.21)}$ \\
& \textbf{N@20} & 0.4780$_{(\color{teal}\downarrow 11.04)}$ & 0.2595$_{(\color{teal}\downarrow 51.70)}$ & 0.5253$_{(\color{teal}\downarrow 2.23)}$ & 0.5035$_{(\color{teal}\downarrow 6.29)}$ & 0.4656$_{(\color{teal}\downarrow 13.34)}$ & 0.4936$_{(\color{teal}\downarrow 8.13)}$ \\
\midrule
\multirow{2}{*}{\textbf{Food}} 
& \textbf{R@20} & 0.7871$_{(\color{teal}\downarrow 6.95)}$ & 0.3732$_{(\color{teal}\downarrow 55.88)}$ & 0.8116$_{(\color{teal}\downarrow 4.05)}$ & 0.8184$_{(\color{teal}\downarrow 3.25)}$ & 0.7951$_{(\color{teal}\downarrow 6.01)}$ & 0.8112$_{(\color{teal}\downarrow 4.10)}$ \\
& \textbf{N@20} & 0.4815$_{(\color{teal}\downarrow 8.15)}$ & 0.2788$_{(\color{teal}\downarrow 46.81)}$ & 0.5122$_{(\color{teal}\downarrow 2.29)}$ & 0.5104$_{(\color{teal}\downarrow 2.63)}$ & 0.4966$_{(\color{teal}\downarrow 5.27)}$ & 0.4978$_{(\color{teal}\downarrow 5.04)}$ \\
\bottomrule
\end{tabular}
}
\caption{Ablation study on different vital components of RaMen. The figures in subscription with the symbol $\color{teal}\downarrow$ denote the \% reduction of performance when each component of the proposed model is omitted.}
\label{tab:ablation}
\end{table*}

\vspace{-4mm}
\subsection{Ablation Study}
\label{sec:ablation}

\subsubsection{Effect of different important components.} 
This study systematically removes the Characteristic Strategy Encoder (\textit{w/o~CrSE}), Collaborative Strategy Encoder (\textit{w/o~CbSE}), Implicit Strategy-aware Learning \textit{(w/o~ISL}), Multi-strategy Alignment \& Discrimination (\textit{w/o~MAD}), textual (\textit{w/o T}), and visual (\textit{w/o~V}) features to investigate the impact of RaMen's core components in optimizing bundle construction strategies. As shown in Table~\ref{tab:ablation}, the findings underscore the significance of learning explicit strategies. Notably, \textit{w/o CrSE} causes a more significant performance degradation than \textit{w/o CbSE} on the sparse dataset POG, revealing the essence of capturing intrinsic information when extrinsic information is limited. Conversely, denser datasets like Electronics and Food show substantial drops in performance \textit{w/o CbSE}, emphasizing the importance of understanding item relations in informed bundling decisions.

In addition, removing ISL leads to notable performance declines, particularly on POG, where detailed item descriptions promote targeting specific fashion segments. This highlights the effectiveness of our intuition in modeling latent shared attributes among items.
These observations also explain the significant effect of textual/visual features on POG compared to other datasets.
Fundamentally, the model derives most of its critical information from the two encoder mechanisms employed in Explicit Strategy-aware Learning (ESL). Meanwhile, ISL serves as a complementary module, which enables the system to construct more productive bundles by enriching item and bundle representations through alignment with latent shared attributes.
Besides, the experiments of '\textit{Only ISL}' were also conducted and yielded extremely low results. This detection is expected, as this setting merely operates on primitive semantic embeddings of items combined with randomly initialized IDs. Completely omitting ESL prevents the model from capturing essential aspects of bundle construction strategies, such as distinctive characteristics and interdependence among items, which makes bundle representation almost meaningless. 
Thus, the inclusion of '\textit{Only ISL}' evaluation is deemed unnecessary, as its modest impact can be inferred from the performance drop observed when ISL is omitted.
Finally, the considerable decreases in performance \textit{w/o MAD} demonstrate that RaMen’s multi-strategy learning architecture, enhanced by transferring supervision signals through MAD module, enables a more comprehensive and robust grasp of product bundling.

\begin{figure}[t!]
    \centering
    \includegraphics[width=1\columnwidth]{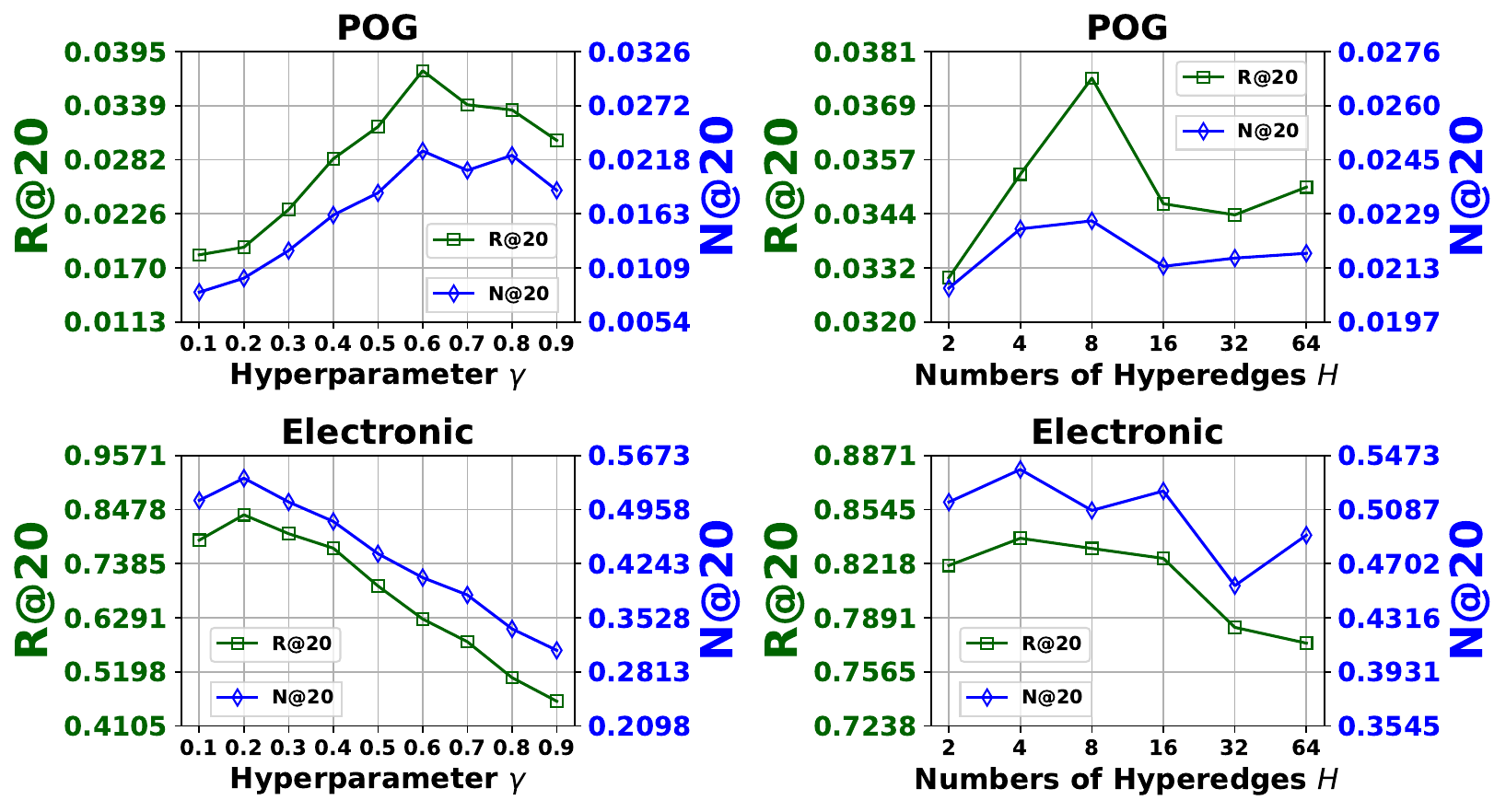}
    \vspace*{-4mm}
    \caption{Impact of $\gamma, H$ on RaMen's performance.}
    \label{fig:hyperparam}
    
\end{figure}

\vspace{-4mm}
\subsubsection{Impacts of critical hyper-parameters.} 
In practice, each domain (\textit{music, fashion, etc.}) has different priorities in weighting strategies for producing bundles. In this work, empirical studies for hyper-parameters are referred to in the related studies where we modeled each component, and tuned it using grid search. These observations are clarified with the aim of providing the essential value range and insights for extending further work. 

Figure \ref{fig:hyperparam} shows the impact of key hyperparameters on the model's performance, specifically the control parameter $\gamma$ in Explicit Strategy-aware Learning and the number of hyperedges $H$ in Implicit Strategy-aware Learning \textit{w.r.t} $R@20$ and $N@20$. The findings demonstrate that RaMen's performance is highly sensitive to $\gamma$ and $H$, requiring careful tuning to achieve optimal performance. 
RaMen can adapt the effect of collaborative strategy (\textit{directly susceptible to data sparsity}) and characteristic strategy through $\gamma$. In essence, the sparser the data, the more crucial the characteristic encoder become.
In particular, the fluctuation of $\gamma$ further reinforces the critical role of both CbSE and CrSE on different data domains, as discussed. As $\gamma$ increases, the performance of RaMen initially improves, reaches an optimal value, and then declines across both datasets. This behavior can be attributed to the role of $\gamma$ as a balancing parameter, which regulates the relative contributions of multi-modal features and collaborative signals in the Explicit Strategy-Aware Learning module. An improperly tuned value of $\gamma$ may disrupt this balance, causing one encoder to dominate the other, ultimately leading to suboptimal performance. 
Regarding the hyperparameter $H$, both small and large-scale datasets benefit from a relatively low number of hyperedges ($4$ or $8$), which conserves memory for computational resources while maintaining competitive performance.

\begin{figure}[t]
    \centering
    \includegraphics[width=1\columnwidth]{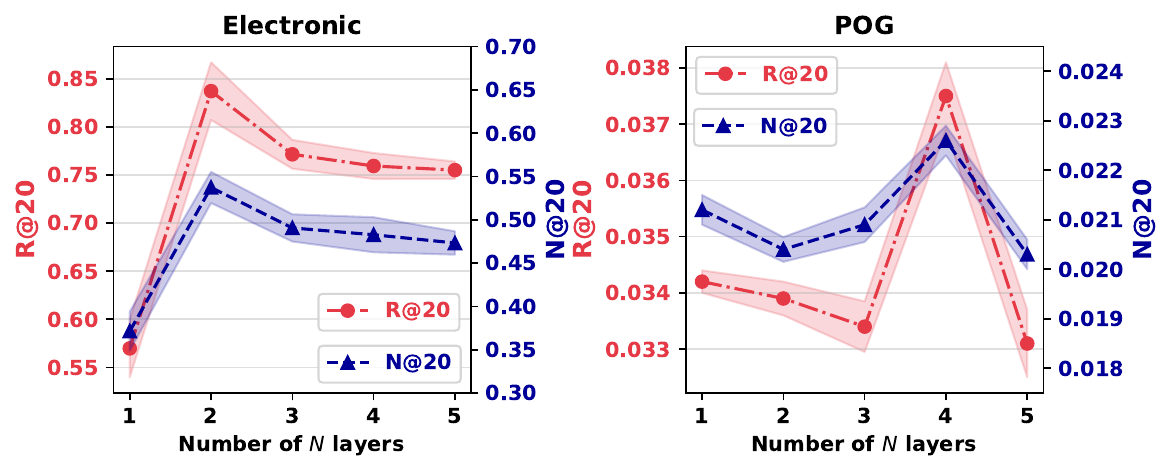}
    \vspace*{-4mm}
    \caption{Impact of hyperparameter $N$ on RaMen's performance.}
    \label{fig:layer_gat_tuning}
\end{figure}

The figure~\ref{fig:layer_gat_tuning} illustrates the impact of varying the number of our refined attention layers ($N$ layers) in the item-item graph component of ESL on RaMen’s performance, evaluated on $R@20$ and $N@20$.
Two datasets (Electronic and POG) are shown to demonstrate the diversity in data size and sparsity. Besides, similar observations are obtained on other datasets.
As shown in figure~\ref{fig:layer_gat_tuning}, the gain in the sparser dataset POG is more modest. The optimal number of attention layers varies: $2$ for Electronic, $4$ for POG. 
This indicates that the structure and density of the item-item graph significantly influence how deep the network should be.
In both datasets, stacking too many layers (e.g., value of $5$) degrades performance, due to oversmoothing - a common issue in GNNs where node representations become indistinguishably similar.
More layers allow for the capture of higher-order relationships as well as multi-context interdependence among items in the bundle construction problem, but they also increase the risk of propagating irrelevant or redundant information, especially in large or noisy graphs. 
The oversmoothing problem is also easily obtained with hypergraphs via the observation of $H$.


\begin{figure}[t]
    \centering
    \includegraphics[width=1\columnwidth]{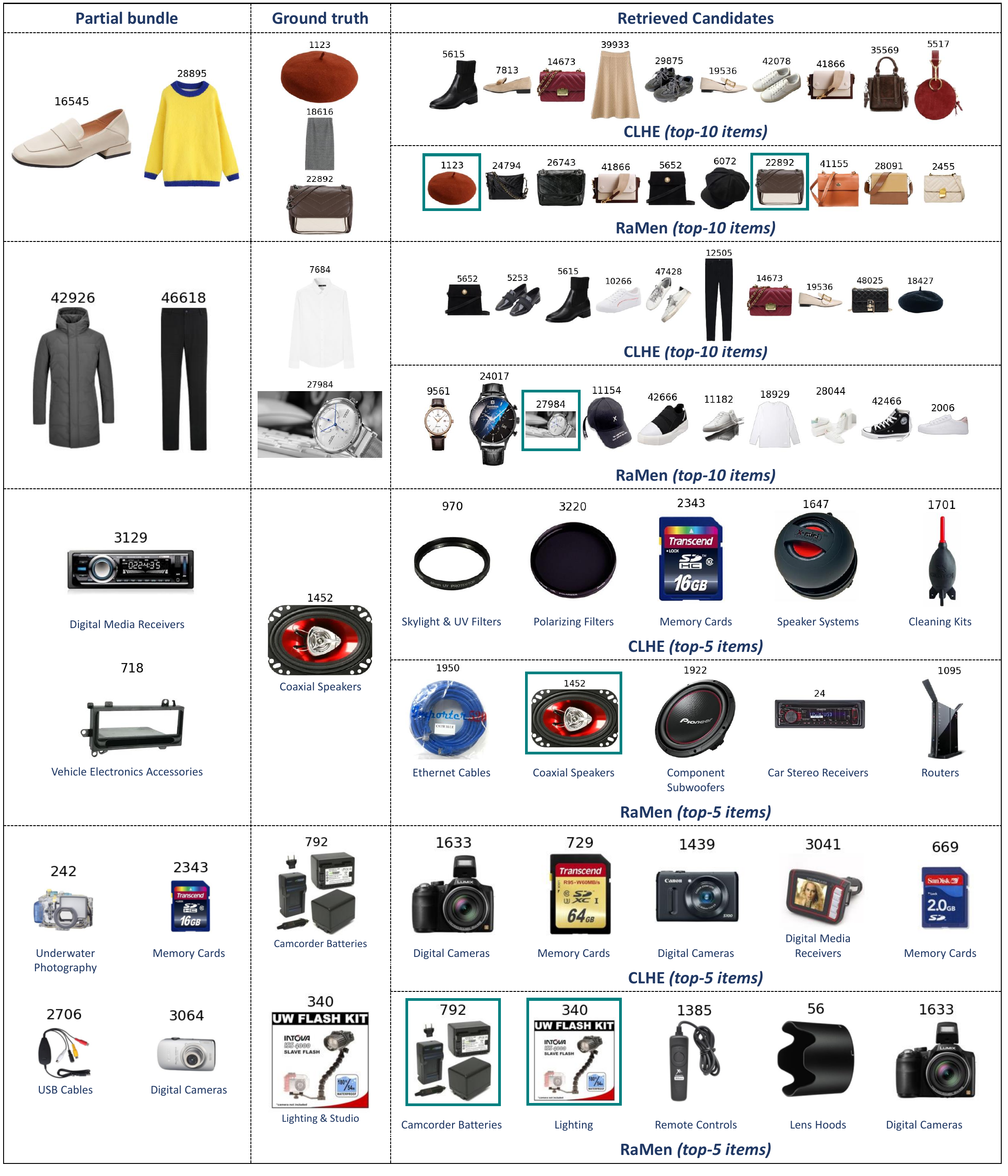}
    \vspace*{-4mm}
    \caption{Practical studies of top-$K$ item candidates of RaMen compared to CLHE across the POG and Electronic dataset. The $\color{teal}\textbf{green box}$ surrounding the item's image denotes the item that the model has correctly predicted.}
    \label{fig:showcase}
\end{figure}

\vspace{-4mm}
\subsection{Qualitative Showcases}

The results validate the effectiveness of multi-strategy multi-modal learning for automatic bundle construction, particularly in practical scenarios like bundles in POG (bundles with IDs as $2234$ and $18762$) and Electronic (bundles with IDs as $507$ and $1090$), which is aimlessly selected among sets of prominent predicted cases and depicted in Figure~\ref{fig:showcase}. In both scenarios, RaMen demonstrates superior capability in capturing bundling strategies compared to the state-of-the-art model CLHE, particularly in retrieving complementary items aligned with user intents. In contrast, CLHE primarily relies on semantic information and fails to exploit complex relations among items, resulting in not only biases towards substitute products (e.g., \textit{memory cards}, \textit{cameras} in Electronics; and \textit{pants} in POG), but also the misconception of product segmentation (e.g., POG bundle aimed at \textit{men's fashion}). 
Notably, in the first example, $5$ out of the $10$ items predicted by CLHE are shoes, which, although similar to the ground truth, reveal a tendency of CLHE to over-rely on multi-modal feature similarity. This approach often results in the prediction of alternative rather than complementary items, a limitation in cases such as fashion bundles, where a balanced selection of items is crucial.
In the third example, most of the items predicted by RaMen belong to the "\textit{Car \& Vehicle Electronics}" category, aligning well with the partial bundle's intent. In contrast, none of CLHE’s predictions are relevant to the bundle’s intended purpose. This highlights RaMen’s superior ability to grasp bundle intents, a strength likely driven by the hypergraph structure in its Implicit Strategy-aware Learning module, where items with related information may be strongly connected with the same hyperedges.
By explicitly leveraging item characteristics and collaborations, and integrating implicit shared attributes, RaMen overcomes these shortcomings of SOTA models. Moreover, these observations provide valuable insights into how multi-strategy approaches can further improve performance in the recommendation of complex item sets~\cite{li2023next,sun2024revisiting,ma2024leveraging,sun2024survey}.

\vspace{-4mm}
\section{Conclusion}
This study introduces RaMen, a novel bundle construction approach that effectively integrates intrinsic and extrinsic information using Explicit and Implicit Strategy-aware Learning. Through extensive experiments across multiple domains and datasets, RaMen consistently outperforms state-of-the-art methods. Moreover, RaMen demonstrates robustness in handling noise within large bundle structures by effectively modeling distinct decision-making strategies and facilitating knowledge transfer between them. Experiment analyses demystify RaMen’s ability to model and integrate diverse decision-making strategies, providing a comprehensive and robust framework for solving bundle construction. Our insightful illustrations can further explore handling even larger bundle structures and refine strategy alignment techniques for complex item set recommendations.

\vspace{-4mm}

\newpage


\bibliography{mybibfile}

\begin{thebibliography}{43}
\providecommand{\natexlab}[1]{#1}
\providecommand{\url}[1]{\texttt{#1}}
\expandafter\ifx\csname urlstyle\endcsname\relax
  \providecommand{\doi}[1]{doi: #1}\else
  \providecommand{\doi}{doi: \begingroup \urlstyle{rm}\Url}\fi

\bibitem[Brody et~al.(2022)Brody, Alon, and Yahav]{brodyattentive}
S.~Brody, U.~Alon, and E.~Yahav.
\newblock How attentive are graph attention networks?
\newblock In \emph{International Conference on Learning Representations}, 2022.

\bibitem[Bui et~al.(2024)Bui, Nguyen, Thi, Le, and Le]{bui2024bridge}
T.-N. Bui, H.-S. Nguyen, C.-V.~N. Thi, H.-Q. Le, and D.-T. Le.
\newblock Bridge: Bundle recommendation via instruction-driven generation.
\newblock \emph{arXiv preprint arXiv:2412.18092}, 2024.

\bibitem[Bui et~al.(2025)Bui, Nguyen, Nguyen, Le, and Le]{bui2025personalized}
T.-N. Bui, H.-S. Nguyen, C.-V.~T. Nguyen, H.-Q. Le, and D.-T. Le.
\newblock Personalized diffusion model reshapes cold-start bundle recommendation.
\newblock In \emph{Companion Proceedings of the ACM on Web Conference 2025}, pages 3088--3091, 2025.

\bibitem[Chang et~al.(2021)Chang, Gao, He, Jin, and Li]{chang2021bundle}
J.~Chang, C.~Gao, X.~He, D.~Jin, and Y.~Li.
\newblock Bundle recommendation and generation with graph neural networks.
\newblock \emph{IEEE Transactions on Knowledge and Data Engineering}, 35\penalty0 (3):\penalty0 2326--2340, 2021.

\bibitem[Chen et~al.(2018)Chen, Lamere, Schedl, and Zamani]{chen2018recsys}
C.-W. Chen, P.~Lamere, M.~Schedl, and H.~Zamani.
\newblock Recsys challenge 2018: Automatic music playlist continuation.
\newblock In \emph{Proceedings of the 12th ACM Conference on Recommender Systems}, pages 527--528, 2018.

\bibitem[Chen et~al.(2019{\natexlab{a}})Chen, Liu, He, Gao, and Zheng]{chen2019matching}
L.~Chen, Y.~Liu, X.~He, L.~Gao, and Z.~Zheng.
\newblock Matching user with item set: Collaborative bundle recommendation with deep attention network.
\newblock In \emph{IJCAI}, pages 2095--2101, 2019{\natexlab{a}}.

\bibitem[Chen et~al.(2019{\natexlab{b}})Chen, Huang, Xu, Guo, Guo, Sun, Li, Pfadler, Zhao, and Zhao]{chen2019pog}
W.~Chen, P.~Huang, J.~Xu, X.~Guo, C.~Guo, F.~Sun, C.~Li, A.~Pfadler, H.~Zhao, and B.~Zhao.
\newblock Pog: personalized outfit generation for fashion recommendation at alibaba ifashion.
\newblock In \emph{Proceedings of the 25th ACM SIGKDD international conference on knowledge discovery \& data mining}, pages 2662--2670, 2019{\natexlab{b}}.

\bibitem[Deng et~al.(2021)Deng, Wang, Zhao, Wu, Ding, Zou, Shang, Tao, and Fan]{deng2021build}
Q.~Deng, K.~Wang, M.~Zhao, R.~Wu, Y.~Ding, Z.~Zou, Y.~Shang, J.~Tao, and C.~Fan.
\newblock Build your own bundle-a neural combinatorial optimization method.
\newblock In \emph{Proceedings of the 29th ACM International Conference on Multimedia}, pages 2625--2633, 2021.

\bibitem[Deng et~al.(2023)Deng, Li, Guo, Liu, Zou, and Li]{deng2023multi}
Z.~Deng, J.~Li, Z.~Guo, W.~Liu, L.~Zou, and G.~Li.
\newblock Multi-view multi-aspect neural networks for next-basket recommendation.
\newblock In \emph{Proceedings of the 46th International ACM SIGIR Conference on Research and Development in Information Retrieval}, pages 1283--1292, 2023.

\bibitem[Ding et~al.(2023)Ding, Mok, Ma, and Bin]{ding2023personalized}
Y.~Ding, P.~Mok, Y.~Ma, and Y.~Bin.
\newblock Personalized fashion outfit generation with user coordination preference learning.
\newblock \emph{Information Processing \& Management}, 60\penalty0 (5):\penalty0 103434, 2023.

\bibitem[Du et~al.(2023)Du, Qian, Ma, and Xiang]{du2023enhancing}
X.~Du, K.~Qian, Y.~Ma, and X.~Xiang.
\newblock Enhancing item-level bundle representation for bundle recommendation.
\newblock \emph{ACM Transactions on Recommender Systems}, 2023.

\bibitem[Fang et~al.(2018)Fang, Xiao, Wang, and Lan]{fang2018customized}
Y.~Fang, X.~Xiao, X.~Wang, and H.~Lan.
\newblock Customized bundle recommendation by association rules of product categories for online supermarkets.
\newblock In \emph{2018 IEEE Third International Conference on Data Science in Cyberspace (DSC)}, pages 472--475. IEEE, 2018.

\bibitem[Gao et~al.(2023)Gao, Zheng, Li, Li, Qin, Piao, Quan, Chang, Jin, He, et~al.]{gao2023survey}
C.~Gao, Y.~Zheng, N.~Li, Y.~Li, Y.~Qin, J.~Piao, Y.~Quan, J.~Chang, D.~Jin, X.~He, et~al.
\newblock A survey of graph neural networks for recommender systems: Challenges, methods, and directions.
\newblock \emph{ACM Transactions on Recommender Systems}, 1\penalty0 (1):\penalty0 1--51, 2023.

\bibitem[Guo et~al.(2024)Guo, Li, Li, Wang, Shi, and Ruan]{guo2024lgmrec}
Z.~Guo, J.~Li, G.~Li, C.~Wang, S.~Shi, and B.~Ruan.
\newblock Lgmrec: Local and global graph learning for multimodal recommendation.
\newblock In \emph{Proceedings of the AAAI Conference on Artificial Intelligence}, volume~38, pages 8454--8462, 2024.

\bibitem[Han et~al.(2017)Han, Wu, Jiang, and Davis]{han2017learning}
X.~Han, Z.~Wu, Y.-G. Jiang, and L.~S. Davis.
\newblock Learning fashion compatibility with bidirectional lstms.
\newblock In \emph{Proceedings of the 25th ACM international conference on Multimedia}, pages 1078--1086, 2017.

\bibitem[He et~al.(2020)He, Deng, Wang, Li, Zhang, and Wang]{he2020lightgcn}
X.~He, K.~Deng, X.~Wang, Y.~Li, Y.~Zhang, and M.~Wang.
\newblock Lightgcn: Simplifying and powering graph convolution network for recommendation.
\newblock In \emph{Proceedings of the 43rd International ACM SIGIR conference on research and development in Information Retrieval}, pages 639--648, 2020.

\bibitem[Irene et~al.(2019)Irene, Borrelli, Zanoni, Buccoli, and Sarti]{irene2019automatic}
R.~T. Irene, C.~Borrelli, M.~Zanoni, M.~Buccoli, and A.~Sarti.
\newblock Automatic playlist generation using convolutional neural networks and recurrent neural networks.
\newblock In \emph{2019 27th European signal processing conference (EUSIPCO)}, pages 1--5. IEEE, 2019.

\bibitem[Jang et~al.(2017)Jang, Gu, and Poole]{jang2017categorical}
E.~Jang, S.~Gu, and B.~Poole.
\newblock Categorical reparametrization with gumbel-softmax.
\newblock In \emph{International Conference on Learning Representations (ICLR 2017)}. OpenReview. net, 2017.

\bibitem[Kingma and Ba(2015)]{kingma2014adam}
D.~P. Kingma and J.~Ba.
\newblock Adam: {A} method for stochastic optimization.
\newblock In Y.~Bengio and Y.~LeCun, editors, \emph{3rd International Conference on Learning Representations}, 2015.

\bibitem[Li et~al.(2022)Li, Li, Xiong, and Hoi]{li2022blip}
J.~Li, D.~Li, C.~Xiong, and S.~Hoi.
\newblock Blip: Bootstrapping language-image pre-training for unified vision-language understanding and generation.
\newblock In \emph{International conference on machine learning}, pages 12888--12900. PMLR, 2022.

\bibitem[Li et~al.(2023)Li, Jullien, Ariannezhad, and de~Rijke]{li2023next}
M.~Li, S.~Jullien, M.~Ariannezhad, and M.~de~Rijke.
\newblock A next basket recommendation reality check.
\newblock \emph{ACM Transactions on Information Systems}, 41\penalty0 (4):\penalty0 1--29, 2023.

\bibitem[Liu et~al.(2022)Liu, Chen, Cheng, Liu, Nie, and Kankanhalli]{liu2022disentangled}
F.~Liu, H.~Chen, Z.~Cheng, A.~Liu, L.~Nie, and M.~Kankanhalli.
\newblock Disentangled multimodal representation learning for recommendation.
\newblock \emph{IEEE Transactions on Multimedia}, 25:\penalty0 7149--7159, 2022.

\bibitem[Liu et~al.(2017)Liu, Fu, Chen, Xiong, and Chen]{liu2017modeling}
G.~Liu, Y.~Fu, G.~Chen, H.~Xiong, and C.~Chen.
\newblock Modeling buying motives for personalized product bundle recommendation.
\newblock \emph{ACM Transactions on Knowledge Discovery from Data (TKDD)}, 11\penalty0 (3):\penalty0 1--26, 2017.

\bibitem[Liu et~al.(2023)Liu, Hu, Xiao, Zhao, Gao, Wang, Li, and Tang]{liu2023multimodal}
Q.~Liu, J.~Hu, Y.~Xiao, X.~Zhao, J.~Gao, W.~Wang, Q.~Li, and J.~Tang.
\newblock Multimodal recommender systems: A survey.
\newblock \emph{ACM Computing Surveys}, 2023.

\bibitem[Liu et~al.(2025)Liu, Wu, Tao, Ma, Wei, and Chua]{liu2025fine}
X.~Liu, J.~Wu, Z.~Tao, Y.~Ma, Y.~Wei, and T.-s. Chua.
\newblock Fine-tuning multimodal large language models for product bundling.
\newblock In \emph{Proceedings of the 31st ACM SIGKDD Conference on Knowledge Discovery and Data Mining V. 1}, pages 848--858, 2025.

\bibitem[Ma et~al.(2022)Ma, He, Zhang, Wang, and Chua]{ma2022crosscbr}
Y.~Ma, Y.~He, A.~Zhang, X.~Wang, and T.-S. Chua.
\newblock Crosscbr: Cross-view contrastive learning for bundle recommendation.
\newblock In \emph{Proceedings of the 28th ACM SIGKDD Conference on Knowledge Discovery and Data Mining}, pages 1233--1241, 2022.

\bibitem[Ma et~al.(2024{\natexlab{a}})Ma, He, Zhong, Wang, Zimmermann, and Chua]{ma2024cirp}
Y.~Ma, Y.~He, W.~Zhong, X.~Wang, R.~Zimmermann, and T.-S. Chua.
\newblock Cirp: Cross-item relational pre-training for multimodal product bundling.
\newblock In \emph{Proceedings of the 32nd ACM International Conference on Multimedia}, pages 9641--9649, 2024{\natexlab{a}}.

\bibitem[Ma et~al.(2024{\natexlab{b}})Ma, Liu, Wei, Tao, Wang, and Chua]{ma2024leveraging}
Y.~Ma, X.~Liu, Y.~Wei, Z.~Tao, X.~Wang, and T.-S. Chua.
\newblock Leveraging multimodal features and item-level user feedback for bundle construction.
\newblock In \emph{Proceedings of the 17th ACM International Conference on Web Search and Data Mining}, pages 510--519, 2024{\natexlab{b}}.

\bibitem[Nguyen et~al.(2023)Nguyen, Bui, Nguyen, Can, Nguyen, Le, and Le]{nguyen2023hhmc}
H.-S. Nguyen, T.-N. Bui, L.-H. Nguyen, D.-C. Can, C.-V.~T. Nguyen, D.-T. Le, and H.-Q. Le.
\newblock Hhmc: a heterogeneous x homogeneous graph-based network for multimodal cross-selling recommendation.
\newblock In \emph{the 15th International Conference on Knowledge and Systems Engineering}, pages 1--6. IEEE, 2023.

\bibitem[Nguyen et~al.(2024)Nguyen, Bui, Nguyen, Hoang, Thi~Nguyen, Le, and Le]{nguyen2024bundle}
H.-S. Nguyen, T.-N. Bui, L.-H. Nguyen, H.~Hoang, C.-V. Thi~Nguyen, H.-Q. Le, and D.-T. Le.
\newblock Bundle recommendation with item-level causation-enhanced multi-view learning.
\newblock In \emph{Joint European Conference on Machine Learning and Knowledge Discovery in Databases}, pages 324--341. Springer, 2024.

\bibitem[Oord et~al.(2018)Oord, Li, and Vinyals]{oord2018representation}
A.~v.~d. Oord, Y.~Li, and O.~Vinyals.
\newblock Representation learning with contrastive predictive coding.
\newblock \emph{arXiv preprint arXiv:1807.03748}, 2018.

\bibitem[Pathak et~al.(2017)Pathak, Gupta, and McAuley]{pathak2017generating}
A.~Pathak, K.~Gupta, and J.~McAuley.
\newblock Generating and personalizing bundle recommendations on steam.
\newblock In \emph{Proceedings of the 40th International ACM SIGIR Conference on Research and Development in Information Retrieval}, pages 1073--1076, 2017.

\bibitem[Sun et~al.(2024{\natexlab{a}})Sun, Li, Li, Tao, Zhang, Wang, and Huang]{sun2024survey}
M.~Sun, L.~Li, M.~Li, X.~Tao, D.~Zhang, P.~Wang, and J.~X. Huang.
\newblock A survey on bundle recommendation: Methods, applications, and challenges.
\newblock \emph{arXiv preprint arXiv:2411.00341}, 2024{\natexlab{a}}.

\bibitem[Sun et~al.(2024{\natexlab{b}})Sun, Feng, Yang, Fang, Qu, Ong, and Liu]{sun2024revisiting}
Z.~Sun, K.~Feng, J.~Yang, H.~Fang, X.~Qu, Y.-S. Ong, and W.~Liu.
\newblock Revisiting bundle recommendation for intent-aware product bundling.
\newblock \emph{ACM Transactions on Recommender Systems}, 2\penalty0 (3):\penalty0 1--34, 2024{\natexlab{b}}.

\bibitem[Sun et~al.(2024{\natexlab{c}})Sun, Feng, Yang, Qu, Fang, Ong, and Liu]{sun2024adaptive}
Z.~Sun, K.~Feng, J.~Yang, X.~Qu, H.~Fang, Y.-S. Ong, and W.~Liu.
\newblock Adaptive in-context learning with large language models for bundle generation.
\newblock In \emph{Proceedings of the 47th International ACM SIGIR Conference on Research and Development in Information Retrieval}, pages 966--976, 2024{\natexlab{c}}.

\bibitem[Vaswani(2017)]{vaswani2017attention}
A.~Vaswani.
\newblock Attention is all you need.
\newblock \emph{Advances in Neural Information Processing Systems}, 2017.

\bibitem[Wei et~al.(2023)Wei, Liu, Ma, Wang, Nie, and Chua]{wei2023strategy}
Y.~Wei, X.~Liu, Y.~Ma, X.~Wang, L.~Nie, and T.-S. Chua.
\newblock Strategy-aware bundle recommender system.
\newblock In \emph{Proceedings of the 46th International ACM SIGIR Conference on Research and Development in Information Retrieval}, pages 1198--1207, 2023.

\bibitem[Xia et~al.(2022)Xia, Huang, Xu, Zhao, Yin, and Huang]{xia2022hypergraph}
L.~Xia, C.~Huang, Y.~Xu, J.~Zhao, D.~Yin, and J.~Huang.
\newblock Hypergraph contrastive collaborative filtering.
\newblock In \emph{Proceedings of the 45th International ACM SIGIR conference on research and development in information retrieval}, pages 70--79, 2022.

\bibitem[Yu et~al.(2022)Yu, Li, Chen, and Zheng]{yu2022unifying}
Z.~Yu, J.~Li, L.~Chen, and Z.~Zheng.
\newblock Unifying multi-associations through hypergraph for bundle recommendation.
\newblock \emph{Knowledge-Based Systems}, 255:\penalty0 109755, 2022.

\bibitem[Zhao et~al.(2022)Zhao, Wei, Zou, and Mao]{zhao2022multi}
S.~Zhao, W.~Wei, D.~Zou, and X.~Mao.
\newblock Multi-view intent disentangle graph networks for bundle recommendation.
\newblock In \emph{Proceedings of the AAAI Conference on Artificial Intelligence}, volume~36, pages 4379--4387, 2022.

\bibitem[Zhou et~al.(2023)Zhou, Zhou, Zhang, and Shen]{zhou2023enhancing}
H.~Zhou, X.~Zhou, L.~Zhang, and Z.~Shen.
\newblock Enhancing dyadic relations with homogeneous graphs for multimodal recommendation.
\newblock In \emph{ECAI 2023}, pages 3123--3130. IOS Press, 2023.

\bibitem[Zhou and Shen(2023)]{zhou2023tale}
X.~Zhou and Z.~Shen.
\newblock A tale of two graphs: Freezing and denoising graph structures for multimodal recommendation.
\newblock In \emph{Proceedings of ACM International Conference on Multimedia}, pages 935--943, 2023.

\bibitem[Zhu et~al.(2014)Zhu, Harrington, Li, and Tang]{zhu2014bundle}
T.~Zhu, P.~Harrington, J.~Li, and L.~Tang.
\newblock Bundle recommendation in e-commerce.
\newblock In \emph{Proceedings of the 37th international ACM SIGIR conference on Research \& development in information retrieval}, pages 657--666, 2014.

\end{thebibliography}

\end{document}